# NON-ISOTROPIC ANGULAR DISTRIBUTION FOR VERY SHORT-TIME GAMMA-RAY BURSTS

D. B. Cline, C. Matthey, and S. Otwinowski

University of California Los Angeles,
Department of Physics and Astronomy, Box 951447
Los Angeles, California 90095-1547  USA

**Abstract.** While most gamma-ray bursts (GRBs) are now believed to be from cosmological distances, the origin of very short-time GRBs is still not known.  In the past, we have shown that GRBs with time duration (T90) less than 100 ms may form a separate class of GRBs based on the hardness and time distribution of these events.  We have also shown that the ln $N$ – ln $S$ distribution is consistent with the expectation of quasi-Euclidean distribution of sources.  In this paper, we report the study of the angular location of these GRBs showing a strong deviation from isotropy within the Galactic coordinates of +180° < longitude < 90° and -30° < latitude < 30° .  We have studied the rest of the GRBs and do not find a similar deviation.  This further indicates that the very-short GRBs likely form a separate class of GRBs, most probably from sources of Galactic or near solar origin.

## I. INTRODUCTION

Most  GRBs are now considered to come from cosmological distances; of course, this refers mainly to long-duration bursts.  It is well known that there are at least two distinct classes of GRBs:  one with time duration in the 10s of seconds and one with time duration of about one second (Kouveliotou et al. 1993).  Both of these classes display a completely isotropic angular distribution and a $<V/V_{max}>$ much less than 0.5, which is in concert with cosmological sources (Meegan et al. 1996).

Less is known about GRBs with a much shorter time duration (~50 msec on average).  We have carried out an extensive study of these GRBs (Cline et al. 1999, 1997), and have concluded that:

(A)   The hardness distribution is very different from the long-duration GRBs,

(B)    The $<V/V_{max}>$ is consistent with 0.5 or a quasi-Euclidian distribution of sources.

The first results, with limited statistics, were presented in Cline et al. 1999.  In this paper, we present the results of our study of the angular distribution of the very short GRBs and find that this distribution deviates sharply from an isotropic distribution.  We compare this with that of the short GRBs with a time duration of about 1 second; we then speculate on the possible cause of the asymmetry for the very-short GRBs.

## 2.   DIFFERENT CLASSES OF GRBs

The time distribution T90 for all GRBs from the BATSE detector up to May 26, 2000 is shown in Figure 1. We divide the GRBs into three classes according to time duration:  long, $L$
 ($\tau > 1$ s); medium, $M$ (1 s > $\tau$ > 0.1s); and short, $S$ ($\tau$ < 100 ms).  The duration time of T90 is used for all analysis.  In this paper, we will confine the discussion to the $M$ and $S$ classes of GRBs.  Since these events are adjacent in time, it is important to contrast the behavior.

## 2.1. Comparison between *S* and *M* GRBs angular distribution

We assume that the *S* GRBs constitute a separate class of GRBs and fit the time distribution in Fig. 1 with a three-population model. The fit is excellent but does not give significant evidence for a three-population model.

We now turn to the angular distributions of the *S* and *M* GRBs. In Fig. 2A, we show this distribution for the short bursts (46 events in total). We can see directly that this is not an isotropic distribution. To ascertain the significance of the anisotropy, we break up the Galactic map into eight equal probability regions. In Fig. 2B we show the distribution of events in the eight bins; clearly one bin has a large excess.

To contrast the distribution of the *S* GRBs and to test for possible errors in the analysis, we plot the same distributions for the *M* sample in Figs. 3A and 3B. As can be seen from Fig. 3A, this distribution is consistent with isotropy. Fig. 3B shows the same analysis as Fig. 2B, indicating that there are no bins with a statistically significant deviation from the hypothesis of an isotropic distribution. Of the 46 short *S* GRBs, there are 20 in the excess region. We have studied these tables and can find no real differences between the properties of the excess events and those outside of the excess region.

We can make a preliminary conclusion based on Figs. 2A, 2B, and $<V/V_{max}> = 0.52 \pm 0.06$ that the *S* events are likely from Galactic, or possibly more local (solar neighborhood), sources. This is the first convincing evidence of GRBs that are probably at non-cosmological distances.

## 2.2. The $V/V_{max}$ Discussion

To further contrast the GRBs in the *S* and *M* regions, we have calculated $V/V_{max}$ for each event using the $C_p$ values from the BATSE data. In Fig. 4A, we show the distribution of $V/V_{max}$ for the *S* events. As we have previously noted (Cline et al. 1999), this distribution is totally consistent with $<V/V_{max}> = 0.5$ for a local distribution. In contrast, the same distribution for the *M* events shown in Fig. 4B indicates a $<V/V_{max}>$ much less than 0.5 consistent with the same mean values for the *L* (long) events, which is now widely interpreted as being due to the cosmological sources for those GRBs. It is probable that the *M* events are also from cosmological sources; however the *S* events appear to come from local sources. We note that the short bursts are strongly consistent with a $C_p^{-3/2}$ spectrum, indicating a Euclidean source distribution, as was shown previously (D. Cline et al. (1997). In the medium time duration (from 100 ms to 1 s), the ln *N* – ln *S* distribution seems to be non-Euclidean. The $<V/V_{max}>$ for the *S*, *M*, and *L* class of events is, respectively, $0.76 \pm 0.14$, $0.37 \pm 0.03$, $0.36 \pm 0.01$.

## 2.3. Statistical evaluation

To determine the statistical probability for such a deviation, we calculate the Poisson probability distribution for the eight bins with a total of 46 events (see Figure 2B). The probability of observing 20 events in a single bin is $1.6 \times 10^{-5}$. We consider this a very significant deviation from an isotropic distribution.

We also perform the likelihood analysis, testing two hypotheses:

(h1) Poisson distribution with $\lambda = 5.75 = 46/8$

In this case the logorhythmic probability 1n(p) calculated for the experimental sample is $\simeq -3.13$, while the average ln(p) estimated from $10^5$ randomly generated samples using the Poisson distribution with the same $\lambda$ and total number of events is $\simeq -17.76$ with the standard deviation SD $\simeq 1.74$. We conclude that the observed value is about 7.8 SD below the Poisson average. This corresponds to the ~ $1.4 \times 10^{-6}$ chance to observe such a configuration and discards (h1).

(h2) Poisson distribution with extra source in the "anomalous" angular bin

Testing (h2), we first estimate the hypothesis that the underlying distribution for 7 angular bins except the anomalous one is indeed Poisson with $\lambda = 3.714 = 26/7$. The ln(p) of the experimental sample is, in this case, $\simeq$ -15.29, which is less than one SD ( ~ 1.532) from the average for random Poisson samples ( ~ -13.905). That confirms the Poisson hypothesis.

One can get a rough estimate of the parameters characterizing (h2): $\lambda$ ~ 3.7, X ~ 20 - 3.7 =~ 16. The direct minimization of the likelihood for h2 gives $\lambda$ ~ 3.78±0.7, X ~ 15.7±1.5 in good agreement with the previous estimate.

The likelihood minimum calue is $\simeq$ -15.5 (which should be compared with -15.29).

We consider that these results are strong evidencefor acceptance of (h2) and conclude that $S$ GRBs are distributed isotropically with a mean ~ 3.7, but there is an extra source which yields ~ 16 events in the anomalous angular region.

## 3. SOURCES OF SHORT BURSTS

Independent of a direct association of the GRB events with specific Galactic sources, it would be even harder to explain the distributions in Fig. 2A as being due to extragalactic or even cosmological sources. The value of the <$V/V_{max}$> and the location asymmetry would seem to strongly support a Galactic origin of the sources for the $S$ GRBs. It is also clear that this distinction would equally argue for a separate class of $S$ GRBs from the $M$ or $L$ classes shown in Fig. 1.

We believe that a future study of the $S$ GRB population resulting in a possible shorter time distribution will be fruitful. We may only have detected a small fraction of the short bursts (Nemiroff et al. 1998).

The bulk of the very short bursts identified here all have time duration at or below the BATSE 64-ms integration time. We therefore believe that the BATSE trigger is likely an inefficient method of identifying such events; we also believe that many weak bursts may have been missed. There could be as many missed GRBs of 1 ms duration as the number that have been detected at 10 s.

We have studied the nearby star population and found no pattern that fits the distribution found in Fig. 2A. One explanation for the very short bursts that we have offered before – primordial black hole (PBH) evaporation (Cline & Hong 1992; Cline, Sanders, & Hong 1997) – might not naturally give such a distribution; however, we cannot exclude the possibility that the PBHs are clustered in the Galactic plane in a manner such as the one seen in Fig. 2A. We have also looked at a possible Oort cloud explanation (i.e., comets, comets colliding with PBHs, or comets colliding with each other), and there is no obvious association (see Maoz 1993).

In conclusion, we have carefully studied a class of very short-duration GRBs (less than 100 ms) and find that they:
1. have a harder energy spectrum than the bulk of GRBs,
2. have a <$V/V_{max}$> consistent with 1/2, and
3. display a non-isotropic angular distribution.

We believe our results indicate a separate class of GRBs that are most likely from Galactic sources.

## 4. ACKNOWLEDGMENT

The authors wish to thank Lev Okun, Giuseppe Cocconi, Tsvi Piran, Alvaro de Rujula and Nikita Stepanov for very helpful discussions.

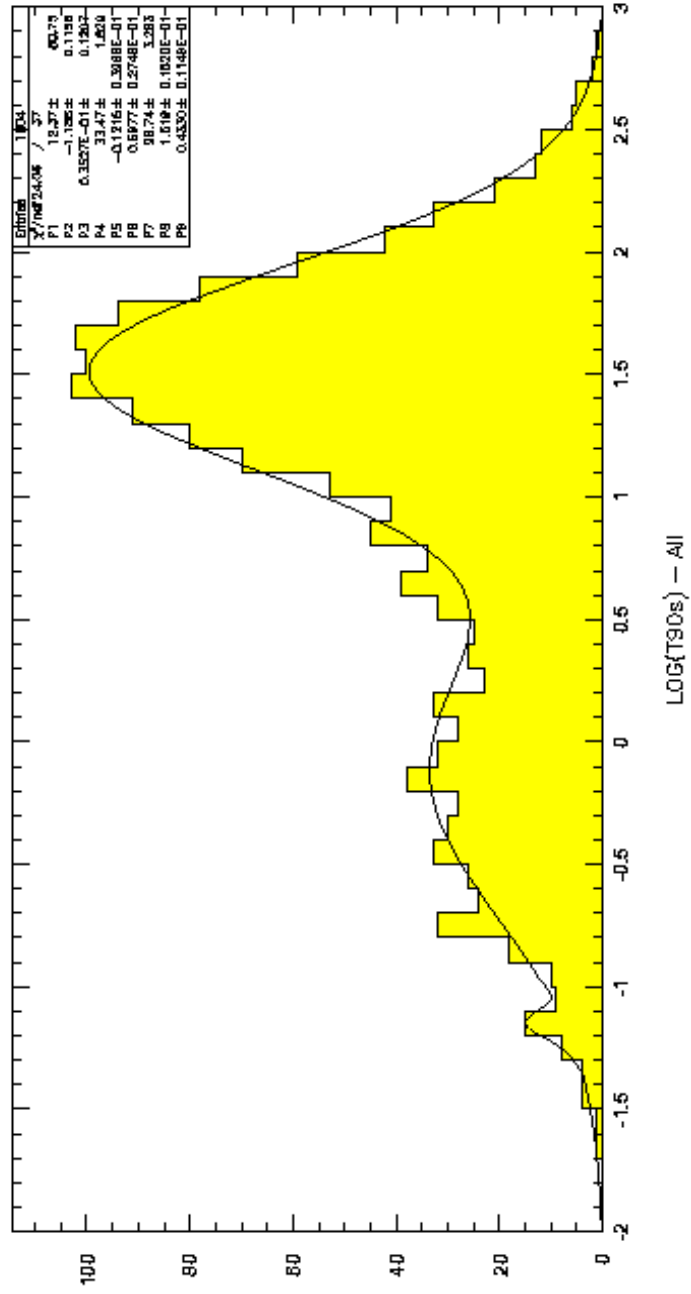

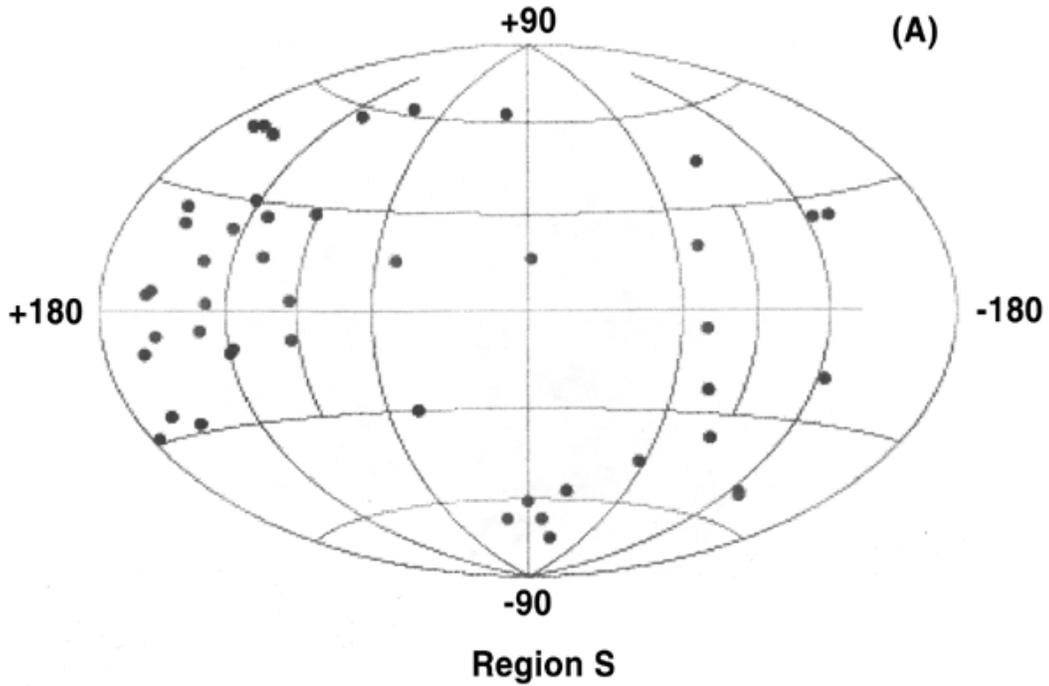

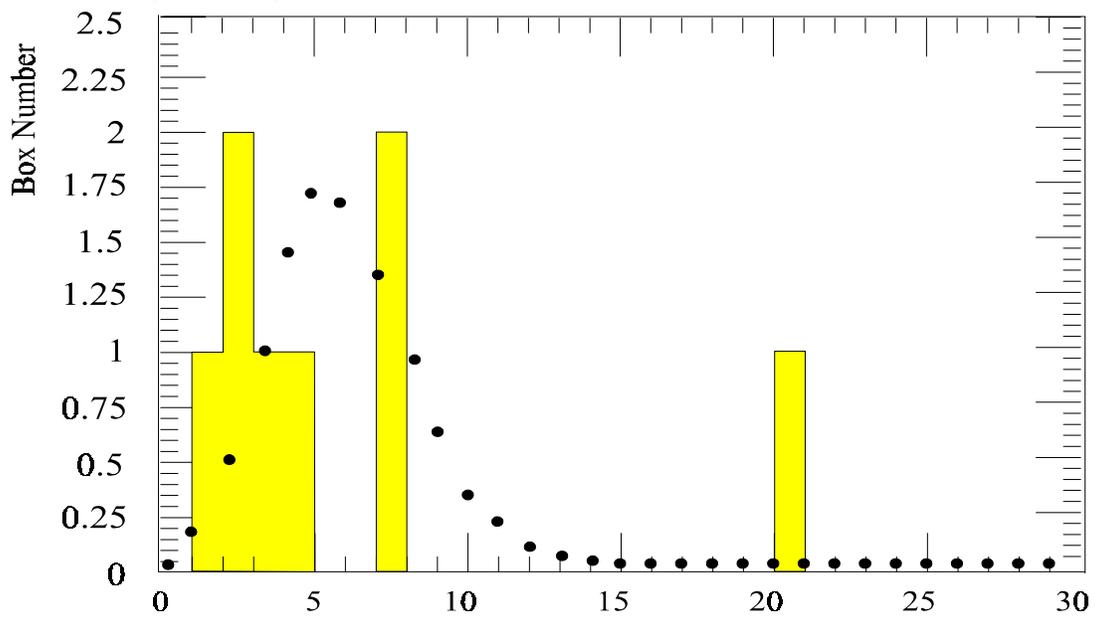

**Figure 2.**

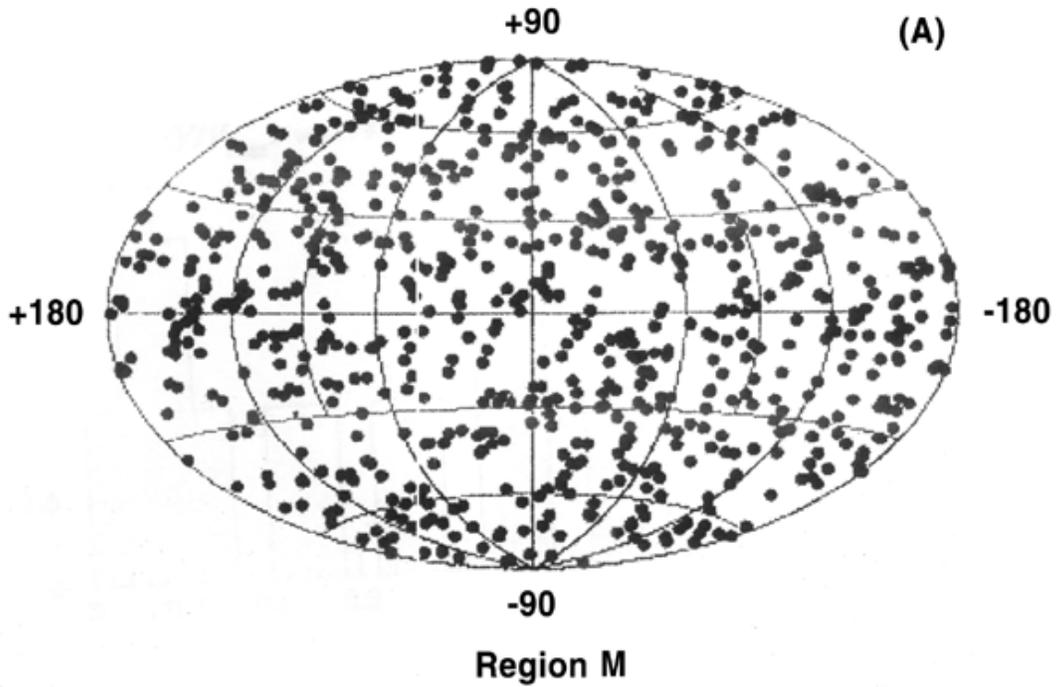

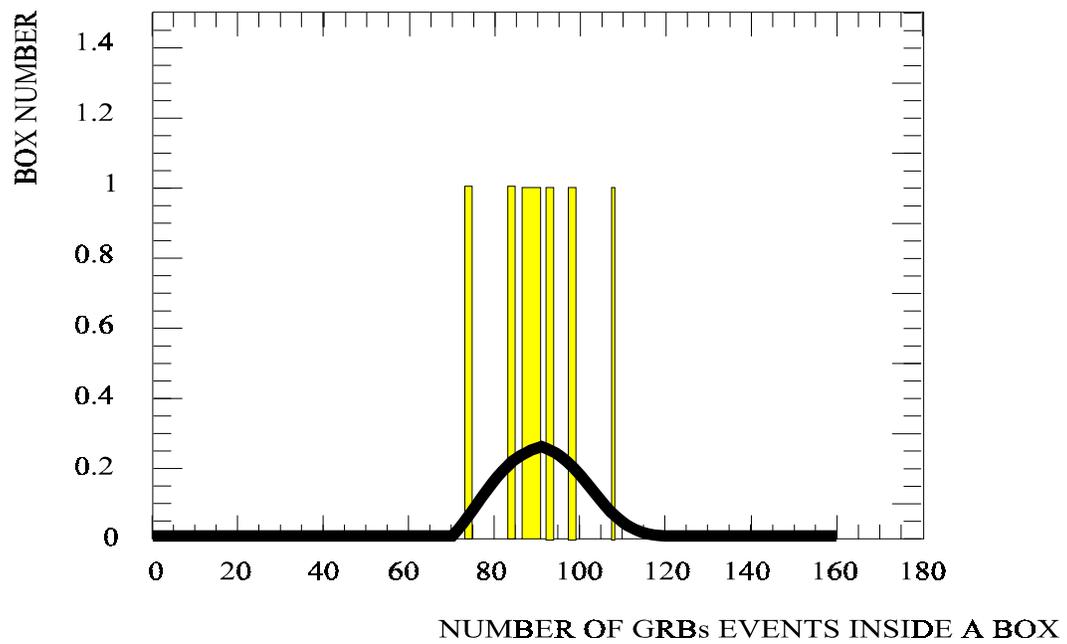

**Figure 3.**

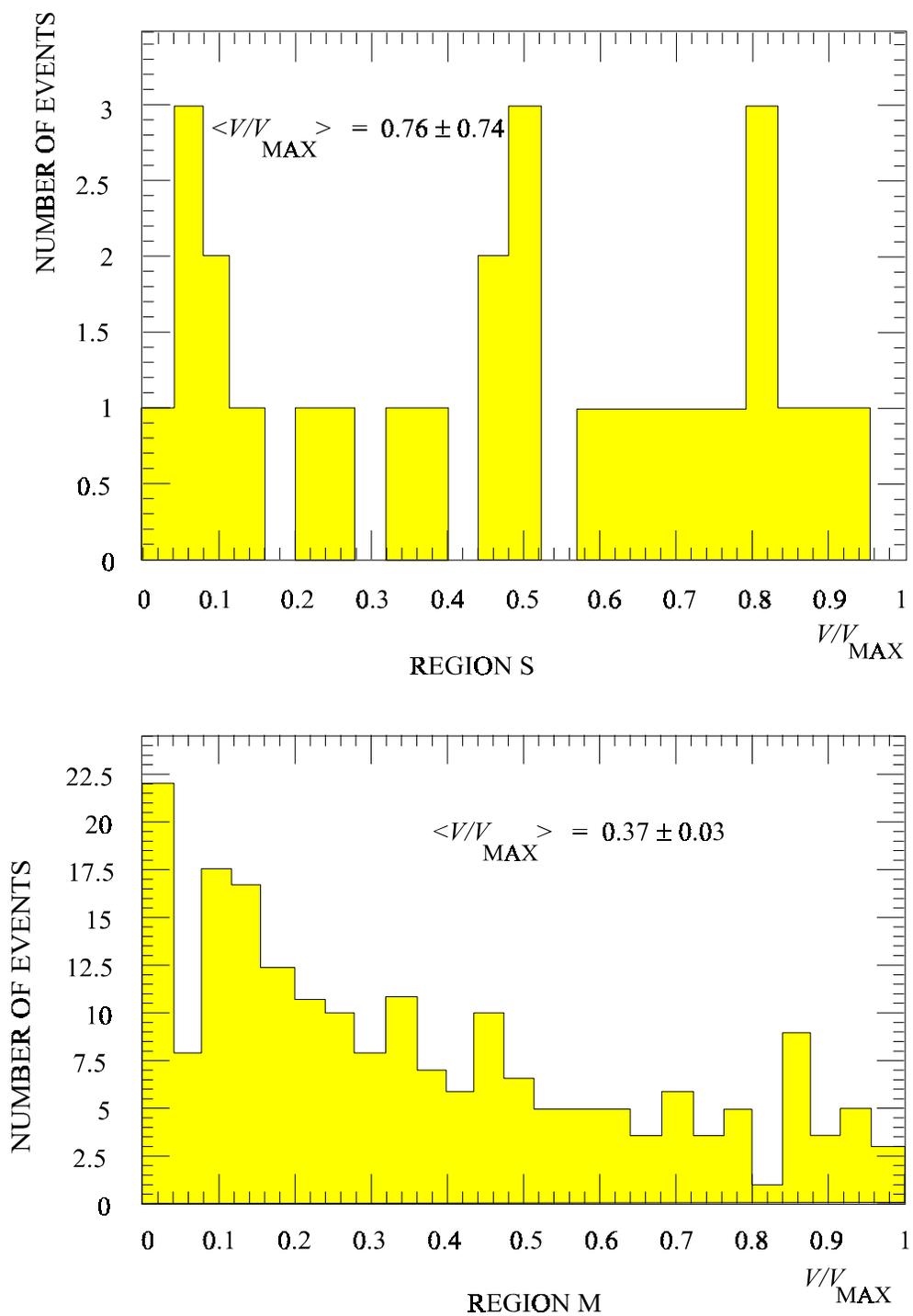

**Figure 4.**